\documentstyle[aps,prl,twocolumn]{revtex}
\begin{document}
\title{
Electric Field Induced Converse Piezoeffect in Screw Dislocated\\
$YBa_2Cu_3O_{7-\delta }$ Thin Films
}
\author{Sergei A. Sergeenkov}
\address{Frank Laboratory of Neutron Physics,
Joint Institute for Nuclear Research,\\ 141980 Dubna, Moscow region, Russia}
\address{
\centering{
\medskip\em
\begin{minipage}{14cm}
{}~~~
A possible scenario for electric-field induced critical
current enhancement in screw dislocated $YBCO$ thin films
is proposed based on the converse piezoelectric effect originated from
a heavily defected medium of the superconducting films in an applied
electric field. The magnitude of the effect is found to depend concurrently
on the film thickness and the number of dislocations inside a superconductor.
{}~\\
\medskip
{}~
\end{minipage}
}}
\maketitle
\narrowtext

Recent experimental results on electric-field induced critical
current enhancement in screw dislocated high-$T_c$ thin films$^{1,2}$ have
been treated as evidence for the field-induced modulation of the mobile
charge carriers density$^{3-5}$. Since high-$T_c$ superconductors (HTS) is
supposed to have a rather low concentration of carriers, one could expect
the field effects to be quite pronounced in these materials. However, even
in extremely thin films (of a few unit cells thick), a rather large
breakdown voltage (of 20-30$V$) is required to produce an essential
modulation of the charge carrier density, and thus to substantially enhance
the critical currents$^{1-5}$. At the same time, the use of the sputtering
technique as well as a laser ablation to prepare thin enough HTS films was
found$^{6-8}$ to bring about a tremendous density of numerous extended
defects (such as screw and edge dislocations) in these thin films. It is
worthwhile to mention that dislocation-induced pinning force enhancement in
zero electric field has been observed earlier practically for the same thin
films$^7$. On the other hand, it is well-known$^{9,10}$ that in ionic
crystals dislocations can accumulate (or trap, due to the electric field
around their cores) electrical charge along their length. Then, application
of an external electric field, in the geometry similar to that used in the
so-called ''superconducting field-effect transistor'' (SuFET) devices$^{1-5}$%
, can sweep up these additional carriers, refreshing the dislocation cores,
that is releasing the dislocation from a cloud of point charges (which, in
field-free configuration, screen an electric field of the dislocation line
for electrical neutrality). Utilizing the ionic (perovskite-like) nature of
the HTS crystals$^{11}$, in the present paper a possibility of the electric
field induced converse piezoeffect and its influence on the critical
currents enhancement in screw dislocated YBCO thin films is discussed.

As is well-known$^{8-10}$, the charges on dislocations play a rather
important role in charge transfer in ionic crystals. If a dislocation is
oriented so that it has an excess of ions of one sign along its core, or if
some ions of predominantly one sign are added to or removed from the end of
half-plane, the dislocation will be charged. In thermal equilibrium it would
be surrounded by a cloud of point defects of the opposite sign to maintain
electrical neutrality. A screw dislocation can transport charge normal to
the Burgers vector if it can carry vacancies with it. Piezoelectric effect,
that is a transverse polarization induced by dislocation motion in ionic
crystals, is certainly cannot be responsible for large field-induced effects
in YBCO thin films because of too low rate of dislocation motion at the
temperatures used in these experiments (another possibility to observe the
direct piezoeffect in dislocated crystals is to apply an external stress
field$^9$). On the contrary, the converse piezoeffect, i.e., a change of
dislocation-induced strain field in applied electric field can produce a
rather considerable change of the critical current density in screw
dislocated HTS thin films. Indeed, according to McElfresh et al.$^8$ (who
treated the correlation of surface topography and flux pinning in YBCO thin
films), the strain field of a dislocation can provide a mechanism by which
the superconducting order parameter can be reduced, making it a possible
site for core pinning. The elastic pinning mechanism could also be
responsible for the large pinning forces associated with dislocation
defects. There are two types of elastic pinning mechanisms possible, a
first-order (parelastic) interaction and a second-order (dielastic)
interaction. In the case of a screw plane defect, the parelastic pinning can
be shown to be negligible$^8$. However, the dielastic interaction comes
about because the self-energy of a defect depends on the elastic constants
of the material in which it forms. Since the crystalline material in a
vortex core is stiffer, there is a higher energy bound up in the defect. For
a tetragonal system with a screw dislocation, the energy density due to the
defect strain is $(1/2)C_{44}\epsilon _{44}^2$, where $C_{44}$ is the shear
modulus and $\epsilon _{44}$ is the shear strain. For a screw plane, $%
\epsilon _{44}=b/2\pi r$, where $r$ is the distance from the center of the
defect and $b$ is the Burgers vector of the defect. The interaction energy
density between the vortex and the screw plane is just $(1/2)\delta
C_{44}\epsilon _{44}^2$, where $\delta C_{44}$ is the difference in shear
modulus between superconducting and normal regions. For a vortex a distance $%
r$ away from the defect, the interaction energy and the pinning force (per
unit length) read, respectively$^8$
\begin{equation}
{\cal E}_d(r)=\frac 12\delta C_{44}\epsilon _{44}^2(\pi \xi ^2)
\end{equation}
and
\begin{equation}
{\cal F}_p\equiv \frac{d{\cal E}_d}{dr}=-\delta C_{44}\left( \frac{b^2\xi ^2
}{4\pi r^3}\right)
\end{equation}
Taking the recently reported for YBCO values of $C_{44}=8\times 10^{10}N/m^2$
and $\delta C_{44}=10^{-5}C_{44}$, setting $r=\xi $ and using a value of $%
b=12\times 10^{-10}m$, the authors of Ref.8 obtained a pinning force per
unit length to be ${\cal F}_p=-10^{-4}N/m$, which corresponds to a critical
current density of $J_c=5\times 10^{10}A/m^2$ using the single vortex limit,
${\cal F}_p=J_c\Phi _0$. These estimates reasonably agree with the $J_c$
found recently in dislocated YBCO thin films$^{1,7}$. Turning to the
field-induced SuFET-type experiments$^{1-5}$, we argue that an applied
electric field can essentially modify the shear strain field of dislocation
\begin{equation}
\epsilon _{44}(\vec E)=\epsilon _{44}(0)+\delta \epsilon _{44}(\vec E),
\end{equation}
where the change of the dislocation-induced strain field in the applied
electric field $\vec E$ is defined as follows$^9$ 
\begin{equation}
\delta \epsilon _{44}(\vec E)=\vec d_{44}\epsilon _0\epsilon _r\vec
E=d_{44}\epsilon _0\epsilon _rE\cos \theta
\end{equation}
Here, $d_{44}$ is the absolute value of the converse piezoeffect
coefficient, $\theta $ stands for the angle between the screw dislocation
line and the direction of an applied electric field, $\epsilon _r$ is the
static permittivity which takes account of the long-range polarization of
the dislocated crystal, and $\epsilon _0=8.85\times 10^{-12}F/m$.

In view of Eqs.(1)-(4), field-induced converse piezoeffect results
in the following changes of the interaction energy ${\cal E}_d$ and the
pinning force density ${\cal F}_p$, respectively 
\begin{equation}
{\cal E}_d(\vec E)=\frac 12\left[ 1+2\frac{\delta \epsilon _{44}(\vec E)}{%
\epsilon _{44}(0)}\right] \delta C_{44}\epsilon _{44}^2(0)(\pi \xi ^2)
\end{equation}
and
\begin{equation}
{\cal F}_p(\vec E)={\cal F}_p(0)\left( 1+\frac{\mid {\vec E}\mid }{E_0}\cos
\theta \right)
\end{equation}
where 
\begin{equation}
E_0=\frac 1{\epsilon _0\epsilon _rd_{44}}
\end{equation}
and ${\cal F}_p(0)$ denotes the pinning force density at zero electric field
(see Eq.(2)). According to Whitworth$^9$, the converse piezoeffect
coefficient $d_{44}$ in ionic dislocated crystals can be presented in the
form 
\begin{equation}
\frac 1{d_{44}}=q\rho b
\end{equation}
Here $\rho $ is the dislocation density,and $q=e^{*}/L$ is the effective
electron charge $e^{*}$ per unit length of dislocation $L$ (in fact, $L$
coincides with the thickness of the YBCO films$^1$).

To attribute the above-mentioned mechanism to the field-induced
changes of the critical current densities in dislocated YBCO thin films, we
propose the following scenario. Depending on the gate polarity, a strong
applied electric field will result either in trapping the additional point
charges by dislocation cores (when $\cos \theta =-1$) reducing the pinning
force density, or in sweeping these point charges (surrounding the
dislocation line to neutralize the net charge) up of the dislocation core
(for the opposite polarity when $\cos \theta =+1$), thus increasing the core
vortex pinning by fresh dislocation line. To make our consideration more
definite, we assume that the total critical current density $J_c$ consists
of two main contributions, the core pinning term $J_{cm}$ which has the
usual form of $J_{cm}=\Phi _0/16\pi \mu _0\xi \lambda ^2$ (see, e.g.$^{7,8}$%
), and the field-induced dielastic pinning by dislocations, $J_{cd}(\vec E)=%
{\cal F}_p(\vec E)/{\Phi _0}$ where the pinning force density ${\cal F}%
_p(\vec E)$ is governed by the Eq.(6). Finally, for the total critical
current density in dislocated thin films we get 
\begin{equation}
J_c(\vec E)=J_c(0)+J_{cd}(0)\frac{\mid {\vec E}\mid }{E_0}\cos \theta ,
\end{equation}
where 
\begin{equation}
J_c(0)=J_{cm}-J_{cd}(0).
\end{equation}
According to the well-known$^{7,8}$ estimates of $J_{cm}$ and $J_{cd}(0)$
for thin YBCO films, we can put $J_{cm}-J_{cd}(0)\simeq J_{cd}(0)$, that is $%
J_c(0)\simeq J_{cd}(0)$. Using the experimental values for the field-free
(at zero gate voltage $V_G$) and field-induced (at gate voltages $V_G=\pm
10V $ corresponding to the applied electric field of $\pm 2.5\times 10^5V/cm$%
) critical current densities obtained in SuFET-type measurements with YBCO
thin films$^1$, namely $J_c(0)=10^8A/m^2,J_c(+10V)=6\times 10^7A/m^2$, and $%
J_c(-10V)=1.2\times 10^8A/m^2$, for an estimate of the threshold field $E_0$
the above Eq.(9) predicts the value of $6\times 10^5V/cm$ which corresponds
to the breakdown voltage $V_0=25V$. Furthermore, using the fact that the
thickness $L$ of YBCO thin films in the experiments carried out by Mannhart
et al.$^1$ was ca. $70\dot A$, for the linear charge per length of
dislocation we get the value $q=e^{*}/L=2\times 10^{-11}C/m$, that is $%
qb\simeq 0.3e^{*}$. This is quite comparable with the typical values known
for the usual ionic dislocated crystals$^{9,10}$. Moreover, in view of
Eqs.(7) and (8), we can estimate the value of the static permittivity, $%
\epsilon _r$, in screw dislocated YBCO thin films. Taking $\rho \cong
10^{10}cm^{-2}$ for the maximum dislocation density observed in these films$%
^{1-3}$, we find $\epsilon _r\cong 0.01$, which reasonably agrees with the
typical permittivities in ionic dislocated crystals$^{9,10}$. Remarkably, as
the threshold field $E_0$ (see Eq.(7)) contains no specifically
superconducting parameters, we can expect that this mechanism will persist
above $T_c$, and will be practically insensitive to applied magnetic fields,
in agreement with the observations$^{1-3}$.

In summary, a possible scenario for electric-field induced critical
current enhancement recently observed in screw dislocated YBCO thin films$^1$
has been proposed based on the converse piezoelectric effect originated from
a heavily defected medium of the superconducting films in an applied
electric field. The magnitude of the effect was found to depend concurrently
on the film thickness and the number of dislocations inside a superconductor.%

\end{document}